\documentclass[twocolumn,floatfix,showpacs,preprintnumbers,amsmath,amssymb,superscriptaddress]{revtex4}

\usepackage{graphicx}  
\usepackage{dcolumn}   
\usepackage{bm}        

\newcommand{\met}{\,/\!\!\!\!E_{T}}

\newcommand{\sht}{H_T}
\newcommand{\ttbar}{{t\bar{t}}}

\newcommand{\tprime}{t^\prime}
\newcommand{\tprimebar}{\bar{t}^\prime}

\begin{document}



\title{Search for Heavy Top-like Quarks 
       Using Lepton Plus Jets Events in
       1.96 TeV $p\bar{p}$ Collisions}

\date{\today}

\begin{abstract}
We present the results of a search for pair production of a new heavy
top-like quark $\tprime$ decaying to a $W$ boson and another quark
using the CDF~II detector in Run~II of the Tevatron $p\bar{p}$
collider.  Using a data sample corresponding to 760~pb$^{-1}$ of
integrated luminosity, we fit the observed spectrum of total
transverse energy and reconstructed $\tprime$ quark mass to a
combination of standard model processes and $\tprime$ pair production.
We see no evidence for $\tprime\tprimebar$ production, and we infer a
lower limit of 256 GeV/$c^2$ on the mass of the $\tprime$ at 95\% CL
assuming standard strong couplings for the $\tprime$.
\end{abstract}


\affiliation{Institute of Physics, Academia Sinica, Taipei, Taiwan 11529, Republic of China} 
\affiliation{Argonne National Laboratory, Argonne, Illinois 60439} 
\affiliation{Institut de Fisica d'Altes Energies, Universitat Autonoma de Barcelona, E-08193, Bellaterra (Barcelona), Spain} 
\affiliation{Baylor University, Waco, Texas  76798} 
\affiliation{Istituto Nazionale di Fisica Nucleare, University of Bologna, I-40127 Bologna, Italy} 
\affiliation{Brandeis University, Waltham, Massachusetts 02254} 
\affiliation{University of California, Davis, Davis, California  95616} 
\affiliation{University of California, Los Angeles, Los Angeles, California  90024} 
\affiliation{University of California, San Diego, La Jolla, California  92093} 
\affiliation{University of California, Santa Barbara, Santa Barbara, California 93106} 
\affiliation{Instituto de Fisica de Cantabria, CSIC-University of Cantabria, 39005 Santander, Spain} 
\affiliation{Carnegie Mellon University, Pittsburgh, PA  15213} 
\affiliation{Enrico Fermi Institute, University of Chicago, Chicago, Illinois 60637} 
\affiliation{Comenius University, 842 48 Bratislava, Slovakia; Institute of Experimental Physics, 040 01 Kosice, Slovakia} 
\affiliation{Joint Institute for Nuclear Research, RU-141980 Dubna, Russia} 
\affiliation{Duke University, Durham, North Carolina  27708} 
\affiliation{Fermi National Accelerator Laboratory, Batavia, Illinois 60510} 
\affiliation{University of Florida, Gainesville, Florida  32611} 
\affiliation{Laboratori Nazionali di Frascati, Istituto Nazionale di Fisica Nucleare, I-00044 Frascati, Italy} 
\affiliation{University of Geneva, CH-1211 Geneva 4, Switzerland} 
\affiliation{Glasgow University, Glasgow G12 8QQ, United Kingdom} 
\affiliation{Harvard University, Cambridge, Massachusetts 02138} 
\affiliation{Division of High Energy Physics, Department of Physics, University of Helsinki and Helsinki Institute of Physics, FIN-00014, Helsinki, Finland} 
\affiliation{University of Illinois, Urbana, Illinois 61801} 
\affiliation{The Johns Hopkins University, Baltimore, Maryland 21218} 
\affiliation{Institut f\"{u}r Experimentelle Kernphysik, Universit\"{a}t Karlsruhe, 76128 Karlsruhe, Germany} 
\affiliation{Center for High Energy Physics: Kyungpook National University, Daegu 702-701, Korea; Seoul National University, Seoul 151-742, Korea; Sungkyunkwan University, Suwon 440-746, Korea; Korea Institute of Science and Technology Information, Daejeon, 305-806, Korea; Chonnam National University, Gwangju, 500-757, Korea} 
\affiliation{Ernest Orlando Lawrence Berkeley National Laboratory, Berkeley, California 94720} 
\affiliation{University of Liverpool, Liverpool L69 7ZE, United Kingdom} 
\affiliation{University College London, London WC1E 6BT, United Kingdom} 
\affiliation{Centro de Investigaciones Energeticas Medioambientales y Tecnologicas, E-28040 Madrid, Spain} 
\affiliation{Massachusetts Institute of Technology, Cambridge, Massachusetts  02139} 
\affiliation{Institute of Particle Physics: McGill University, Montr\'{e}al, Canada H3A~2T8; and University of Toronto, Toronto, Canada M5S~1A7} 
\affiliation{University of Michigan, Ann Arbor, Michigan 48109} 
\affiliation{Michigan State University, East Lansing, Michigan  48824} 
\affiliation{University of New Mexico, Albuquerque, New Mexico 87131} 
\affiliation{Northwestern University, Evanston, Illinois  60208} 
\affiliation{The Ohio State University, Columbus, Ohio  43210} 
\affiliation{Okayama University, Okayama 700-8530, Japan} 
\affiliation{Osaka City University, Osaka 588, Japan} 
\affiliation{University of Oxford, Oxford OX1 3RH, United Kingdom} 
\affiliation{University of Padova, Istituto Nazionale di Fisica Nucleare, Sezione di Padova-Trento, I-35131 Padova, Italy} 
\affiliation{LPNHE, Universite Pierre et Marie Curie/IN2P3-CNRS, UMR7585, Paris, F-75252 France} 
\affiliation{University of Pennsylvania, Philadelphia, Pennsylvania 19104} 
\affiliation{Istituto Nazionale di Fisica Nucleare Pisa, Universities of Pisa, Siena and Scuola Normale Superiore, I-56127 Pisa, Italy} 
\affiliation{University of Pittsburgh, Pittsburgh, Pennsylvania 15260} 
\affiliation{Purdue University, West Lafayette, Indiana 47907} 
\affiliation{University of Rochester, Rochester, New York 14627} 
\affiliation{The Rockefeller University, New York, New York 10021} 
\affiliation{Istituto Nazionale di Fisica Nucleare, Sezione di Roma 1, University of Rome ``La Sapienza," I-00185 Roma, Italy} 
\affiliation{Rutgers University, Piscataway, New Jersey 08855} 
\affiliation{Texas A\&M University, College Station, Texas 77843} 
\affiliation{Istituto Nazionale di Fisica Nucleare, University of Trieste/\ Udine, Italy} 
\affiliation{University of Tsukuba, Tsukuba, Ibaraki 305, Japan} 
\affiliation{Tufts University, Medford, Massachusetts 02155} 
\affiliation{Waseda University, Tokyo 169, Japan} 
\affiliation{Wayne State University, Detroit, Michigan  48201} 
\affiliation{University of Wisconsin, Madison, Wisconsin 53706} 
\affiliation{Yale University, New Haven, Connecticut 06520} 
\author{T.~Aaltonen}
\affiliation{Division of High Energy Physics, Department of Physics, University of Helsinki and Helsinki Institute of Physics, FIN-00014, Helsinki, Finland}
\author{J.~Adelman}
\affiliation{Enrico Fermi Institute, University of Chicago, Chicago, Illinois 60637}
\author{T.~Akimoto}
\affiliation{University of Tsukuba, Tsukuba, Ibaraki 305, Japan}
\author{M.G.~Albrow}
\affiliation{Fermi National Accelerator Laboratory, Batavia, Illinois 60510}
\author{B.~\'{A}lvarez~Gonz\'{a}lez}
\affiliation{Instituto de Fisica de Cantabria, CSIC-University of Cantabria, 39005 Santander, Spain}
\author{S.~Amerio}
\affiliation{University of Padova, Istituto Nazionale di Fisica Nucleare, Sezione di Padova-Trento, I-35131 Padova, Italy}
\author{D.~Amidei}
\affiliation{University of Michigan, Ann Arbor, Michigan 48109}
\author{A.~Anastassov}
\affiliation{Rutgers University, Piscataway, New Jersey 08855}
\author{A.~Annovi}
\affiliation{Laboratori Nazionali di Frascati, Istituto Nazionale di Fisica Nucleare, I-00044 Frascati, Italy}
\author{J.~Antos}
\affiliation{Comenius University, 842 48 Bratislava, Slovakia; Institute of Experimental Physics, 040 01 Kosice, Slovakia}
% ...
\collaboration{CDF Collaboration\footnote{With visitors from $^a$University of Athens, 15784 Athens, Greece, 
$^b$Chinese Academy of Sciences, Beijing 100864, China, 
$^c$University of Bristol, Bristol BS8 1TL, United Kingdom, 
$^d$University Libre de Bruxelles, B-1050 Brussels, Belgium, 
$^e$University of California Irvine, Irvine, CA  92697, 
$^f$University of California Santa Cruz, Santa Cruz, CA  95064, 
$^g$Cornell University, Ithaca, NY  14853, 
$^h$University of Cyprus, Nicosia CY-1678, Cyprus, 
$^i$University College Dublin, Dublin 4, Ireland, 
$^j$University of Edinburgh, Edinburgh EH9 3JZ, United Kingdom, 
$^k$University of Heidelberg, D-69120 Heidelberg, Germany, 
$^l$Universidad Iberoamericana, Mexico D.F., Mexico, 
$^m$University of Manchester, Manchester M13 9PL, England, 
$^n$Nagasaki Institute of Applied Science, Nagasaki, Japan, 
$^o$University de Oviedo, E-33007 Oviedo, Spain, 
$^p$Queen Mary, University of London, London, E1 4NS, England, 
$^q$Texas Tech University, Lubbock, TX  79409, 
$^r$IFIC(CSIC-Universitat de Valencia), 46071 Valencia, Spain, 
}}
\noaffiliation



\maketitle


The discovery of the top quark in 1995~\cite{top} completed the third
generation of fundamental fermions in the quark sector in the standard
model (SM) of particle physics.  A fourth chiral generation of massive
fermions with the same quantum numbers as the known fermions is
predicted in a number of models.  It is favored by flavor
democracy~\cite{democratic}, for example, and arises by unifying spins
and charges in the GUT SO(1,13) framework~\cite{gut}.

Precise measurements from LEP exclude a light fourth neutrino $\nu_4$
with mass $m(\nu_4) < m_Z/2$, where $m_Z$ is the mass of the $Z$
boson.  On the other hand a fourth generation neutrino cannot be too
heavy due to sizeable radiative corrections~\cite{pdg}, although
$m(\nu_4) \approx 100$ GeV/$c^2$ is still consistent with electroweak
data~\cite{kribs}.  If $m(\nu_4) \gtrsim m_Z/2$ the radiative
corrections become small~\cite{maltoni}, such a neutrino may explain
some of the astrophysical puzzles~\cite{astro}, and one extra chiral
family of fermions with quark masses as high as 400 GeV/$c^2$ is
viable~\cite{okun,kribs}.  Additional fermion families can also be
accommodated in two-Higgs-doublet scenarios and $N=2$ SUSY
models~\cite{he}.

In all of the above scenarios the present bounds on the Higgs are
relaxed; the Higgs mass could be as large as 500 GeV/$c^2$ with
enhanced production at the Tevatron and LHC.  In addition, a small
mass splitting between new heavy quarks $\tprime$ and $b^\prime$ is
preferred, such that $m(b^\prime)+m(W) > m(\tprime)$, and $\tprime$
decays predominantly to $Wq$ (a $W$ boson and a down-type quark
$q=d,s,b$)~\cite{frampton}.

Other models with heavy exotic quarks decaying to $Wq$ with vector
couplings to the W boson are possible. Contributions to radiative
corrections from such quarks with mass $M$ decouple as $1/M^2$ and
preserve the agreement with precision data.  For example, the
``beautiful mirrors'' model~\cite{bm} improves the fit to the
precision electroweak observables by eliminating the observed
discrepancy in the $b\bar{b}$ forward-backward asymmetry~\cite{pdg,
chanowitz}.  It introduces a new fermion doublet, a mirror copy of the
standard quark doublets with a heavier version of the SM top decaying
to $Wb$.  

A heavy top-like quark also appears in Little Higgs (LH)
models~\cite{little-higgs}, which evade the hierarchy problem by
introducing a minimal set of gauge and fermion fields in the context
of a large-extra-dimension framework.  In particular, LH models in
which $T$-parity is conserved suggest a massive top-like quark which
can decay to $Wq$, as do LH models requiring two scales ($f_{1,2}$);
these have been shown to prefer a top-like quark having a mass
of approximately 500 GeV~\cite{t-parity,twof}.


In this Letter we present the results of a search for pair production
of a new massive strongly interacting up-type quark $\tprime$ with its
associated antiquark, each decaying to $Wq$, using the large data set
collected by the upgraded Collider Detector at Fermilab (CDF II) in
Run II of the Tevatron.  The data come from $p\bar{p}$ collisions at a
center of mass energy of 1.96 TeV, corresponding to an integrated
luminosity of 760 pb$^{-1}$.

As in the case of $t\bar{t}$ production, the case when one $W$ decays
leptonically leads to events with a lepton, a neutrino, and four
quarks via the chain $t^\prime\tprimebar\to WqWq\to\ell\nu qqqq$.
Employing a selection based on event kinematics avoids imposing a
$b$-quark tagging requirement, which would limit our search to the
decay mode $\tprime\to Wb$.  We select events with a lepton ($e$ or
$\mu$), missing transverse energy~\cite{ref-eta}, and four or more
hadronic jets. 
The observed distributions of the scalar
sum of the transverse energy ($H_T$) of all reconstructed leptons,
jets, and missing transverse energy in these events, together with the
distribution of reconstructed $\tprime$ mass ($M_{rec}$), allow
discrimination of the $\tprime\tprimebar$ signal from the standard
model backgrounds discussed below.


CDF II~\cite{cdf} is a large general purpose detector with an overall
cylindrical geometry surrounding the $p\bar{p}$ interaction region.
The three-dimensional trajectories of charged particles produced in
$p\bar{p}$ collisions are measured 
using multiple layers of silicon microstrip detectors, and at outer
radii with an axial/stereo wire drift chamber.  The tracking system
lies inside a uniform 1.4 T solenoidal magnetic field oriented along 
the beam direction.
Outside the solenoid lie the electromagnetic and hadronic calorimeters, which are
segmented in pseudorapidity ($\eta$) and azimuth in a projective
``tower'' geometry.  
Muons are identified by a system of drift chambers placed
outside the calorimeter steel, which acts as an absorber for hadrons.
The integrated luminosity of the $p\bar{p}$ collisions is measured
using
\v{C}erenkov luminosity counters~\cite{ref-lum}.


Events with a high-$p_T$ (18 GeV/$c$ or more) $e$ or $\mu$ candidate
are identified using high-speed trigger electronics and recorded for
later analysis.  The performance of the trigger and lepton
identification algorithms is described in detail
elsewhere~\cite{top-kin}.  

Jet clustering employs an iterative cone-based technique, which
associates calorimeter energy deposits within a cone of radius $\Delta
R \equiv \sqrt{(\Delta\eta)^2+(\Delta\phi)^2} = 0.4$.  The energies of
reconstructed jets and the missing transverse energy are corrected for
detector non-uniformity and other effects~\cite{jet-energy}.


Selected events must contain an $e$ or $\mu$ having $p_T > 20$
GeV/$c$, four or more jets with $E_T > 15$ GeV and $|\eta|<2.0$, and
missing transverse energy $\met > 20$ GeV.  To ensure that leptons
come from $W$ boson decay they must be isolated; there can be no
significant energy deposit within $\Delta R < 0.4$ of the lepton
momentum.  Also, to ensure that leptons and jets are reconstructed
from the same interaction, the event vertex is required to be within 5
cm of the $z$ position of the lepton track's point of closest approach
to the beam axis.  We observe 451 events in the recorded sample.


The main SM contributions to the selected event sample
come from $t\bar{t}$ events, $W$ plus hadronic jets events, and
hadronic multijet (``QCD'') events having large $\met$ in which one
jet is misreconstructed as a lepton.  We use observed data with
non-isolated leptons to estimate the QCD contribution, following the
same method as in the $t\bar{t}$ cross section measurement~\cite{top-kin}. We
use the \textsc{alpgen}~\cite{alpgen} Monte Carlo generator to simulate 
$W$ plus jets events with \textsc{herwig}~\cite{herwig} used for modeling
parton showers, and the \textsc{pythia}~\cite{pythia} event generator to simulate
both $t\bar{t}$ and $\tprime\tprimebar$ events.  These events pass
through a full detector simulation and reconstruction.

The backgrounds from single top, diboson, and $Z$+jets production 
contribute about 10\% of the accepted events.  However, the kinematic
distributions of interest in this analysis in these processes differ
negligibly from those in $W$+jets events, allowing  use of just the
$W$+jets simulation to model all the non-$t\bar{t}$ background with
real leptons.


For each event we calculate the mass $M_{rec}$ of the hypothetical
$\tprime$ and of the $\tprimebar$ using the same type of kinematic fit
used in a measurement of the top quark mass~\cite{topmass}.  Of all
possible lepton-jet combinations of the four highest-$E_T$ jets,
we select the one with the lowest $\chi^2$ for the hypothesis
$\tprime\to Wq$, having equal reconstructed $\tprime$ and $\tprimebar$
masses, and having the $W$ mass hypothesis satisfied by the relevant jet
pair on one side and by the lepton and $\met$ on the other.  This
procedure selects the correct combination about 30\% of the time.


We perform a binned likelihood fit of background and signal to the
observed two-dimensional distribution of $\sht$ and $M_{rec}$.  The
$\tprime\tprimebar$ events would have larger $\sht$ and $M_{rec}$
than the backgrounds, especially as the $\tprime$ mass gets larger.
Fitting this two-dimensional distribution brings up to 20\% more
sensitivity than fitting either one alone, particularly at lower
$\tprime$ masses. 

Imperfect knowledge of various experimental parameters leads to
systematic uncertainties which degrade our sensitivity to a
$\tprime\tprimebar$ signal.  All systematic effects are represented by
Gaussian-constrained ``nuisance'' parameters in the likelihood.  The
Gaussian width is equal to the systematic uncertainty, except the rate
for $W$+jets-like events, which floats freely in the fit.  We
calculate the likelihood maximized with respect to the nuisance
parameters as a function of a hypothetical $\tprime\tprimebar$ signal
cross section $\sigma$ (assuming a 100\% branching ratio of
$\tprime\to Wq$) and apply Bayes' theorem with a uniform prior in
$\sigma$ to obtain a 95\% C.L. upper limit.

The systematic uncertainty with the largest effect on the final result
is that due to the 3\% uncertainty on the jet energy scale.  The
nuisance parameter representing this effect controls how the
$\sht$-$M_{rec}$ distribution is modified as the jet energy scale
changes within its uncertainty.  We calculate the bin-by-bin
dependence for each background and signal source distribution from
simulated samples in which the jet energy scale has been altered from
its nominal value.


\begin{figure}
  \begin{center}
     \includegraphics[width=\columnwidth]{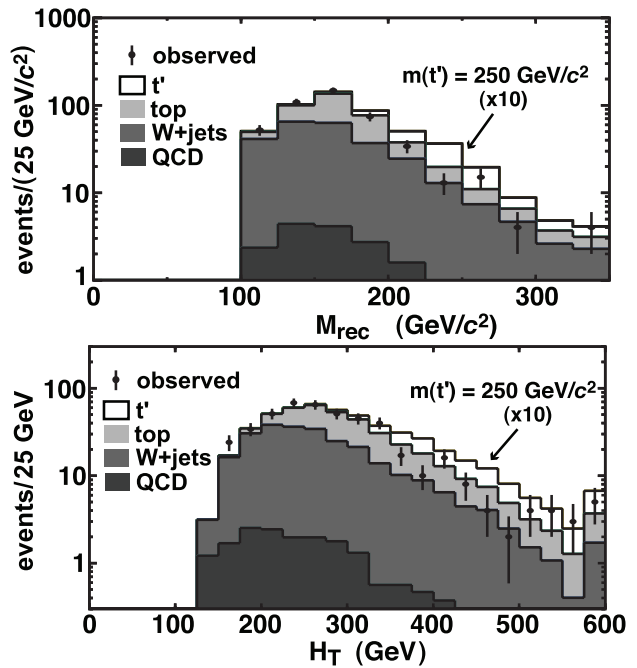}
  \end{center}
  \caption{Observed and predicted distributions of $M_{rec}$ (above) and $\sht$ (below).  The predicted
           distribution corresponds to that for a 250 GeV/$c^2$ mass $\tprime\tprimebar$ 
           signal assuming a value of the cross section at ten times the  
           theoretical one.  Note that in each plot the last bin is an overflow bin,
           and that the $W$+jets contribution also represents other similar backgrounds.}
  \label{fig-Mrec-sht}
\end{figure}

Another systematic uncertainty is due to the lack of knowledge of the
appropriate $Q^2$ scale at which the $W$ plus jets processes should be
evaluated.  The magnitude of this uncertainty comes from changing the
$Q^2$ scale by a factor of two up and down from the nominal choice,
and assigning the larger of the two apparent shifts in $\sigma$ as a
systematic uncertainty.  The effect of this is substantially smaller
than that of the jet energy scale uncertainty.  We also include along
with this effect the uncertainty in the amounts of initial- and
final-state radiation.

Other systematic effects include those due to a 6\% uncertainty in the
integrated luminosity, the 0.7\% uncertainty in lepton identification
efficiencies, and the 27\% uncertainty in QCD background
normalization.  All these have a small effect on the final result.

We constrain the value of the $\ttbar$ production cross section in the
likelihood fit to its theoretical value of 6.7 pb at 175
GeV/$c^2$~\cite{cacciari}.  We assume a 7\% uncertainty in the cross
section, which is predominantly due to uncertainties in the parton
density functions.  We assume that this effect is correlated
positively between the $\tprime\tprimebar$ and $t\bar{t}$ production
processes.


The likelihoods reveal no significant excess attributable to
$\tprime\tprimebar$ production, and in fact the observed distributions
agree well with the zero-signal hypothesis.  Table~\ref{tab-result}
shows the result for the 95\% C.L. upper limit on
$\sigma(p\bar{p}\to\tprime\tprimebar)$ as a function of $\tprime$
mass, assuming that the branching ratio $\tprime\to Wq$ is 100\%.
Figure~\ref{fig-Mrec-sht} shows the observed distributions projected
onto the $M_{rec}$ and $\sht$ dimensions.  The figures compare the
observed distributions with the fit to the background plus a 250
GeV/$c^2$ $\tprime$ signal. 

\begin{figure}
  \begin{center}
     \includegraphics[width=\columnwidth]{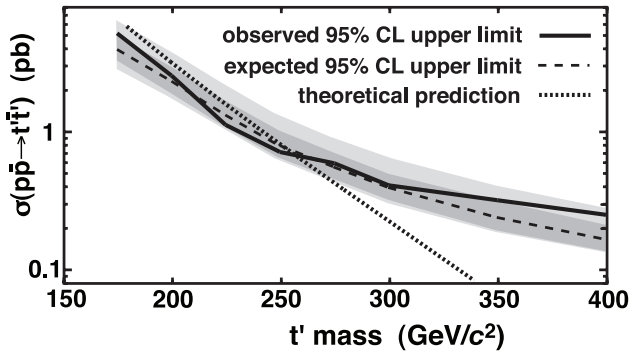}
  \end{center}
  \caption{Observed and expected 95\% C.L. upper limits on the cross section 
           for $\tprime\tprimebar$ production as a function of $\tprime$ mass.  The grey
           bands around the median expected limit show the $\pm$1- and $\pm$2-standard-deviation
           ranges. The theoretical prediction is also shown (assuming a 100\% branching ratio 
           to $Wq$).}
  \label{fig-limit}
\end{figure}

To obtain a lower bound on the mass of the $\tprime$, we compare our
upper limit on $\sigma$ to the theoretical cross section for a
fourth-generation $\tprime$ with SM couplings~\cite{cacciari},
assuming a 100\% branching ratio $B(\tprime\to Wq)$; this is 
illustrated in Figure~\ref{fig-limit}.
We take the point in $\tprime$ mass where the observed limit crosses
the theoretical cross section as the lower bound on the mass of the
$\tprime$: $m(\tprime)$ $>$ 256 GeV/$c^2$, at 95\% C.L.

\begin{table}
\begin{tabular}{lcccccccc}\hline\hline
mass (GeV/$c^2$)  & 175  & 200  & 225  & 250  & 275  & 300  & 350  & 400   \\ \hline
upper limit (pb)  & 5.21 & 2.57 & 1.13 & 0.72 & 0.59 & 0.41 & 0.32 & 0.25  \\ \hline \hline
\end{tabular}
  \caption{The 95\% C.L. upper limits on $\sigma(\tprime\tprimebar)$
           as a function of $\tprime$ mass.}
  \label{tab-result}
\end{table}


In conclusion we present here the result of a search for a new heavy
top-like quark $\tprime$ decaying to $Wq$, using a data sample from
760 pb$^{-1}$ integrated luminosity of $p\bar{p}$ collisions at 1.96
TeV center of mass energy.  Our fit of the observed $\sht$-$M_{rec}$
distribution reveals no excess from $\tprime\tprimebar$ production,
and so we conclude that the mass of the $\tprime$, if it exists, must
exceed 256 GeV/$c^2$ at 95\% C.L. or the $\tprime$ must decay to a
different final state.


We thank the Fermilab staff and the technical staffs of the
participating institutions for their vital contributions. This work
was supported by the U.S. Department of Energy and National Science
Foundation; the Italian Istituto Nazionale di Fisica Nucleare; the
Ministry of Education, Culture, Sports, Science and Technology of
Japan; the Natural Sciences and Engineering Research Council of
Canada; the National Science Council of the Republic of China; the
Swiss National Science Foundation; the A.P. Sloan Foundation; the
Bundesministerium f\"ur Bildung und Forschung, Germany; the Korean
Science and Engineering Foundation and the Korean Research Foundation;
the Science and Technology Facilities Council and the Royal Society,
UK; the Institut National de Physique Nucleaire et Physique des
Particules/CNRS; the Russian Foundation for Basic Research; the
Comisi\'on Interministerial de Ciencia y Tecnolog\'{\i}a, Spain; the
European Community's Human Potential Programme; the Slovak R\&D Agency;
and the Academy of Finland.



\begin{thebibliography}{99}

   \bibitem{top}         F. Abe {\em et al.} (CDF Collaboration), Phys. Rev. Lett.  {\bf 74}, 2626 (1995);
                         S. Abachi {\em et al.} (D0 Collaboration), Phys. Rev. Lett. {\bf 74}, 2632 (1995).

   \bibitem{democratic}  J. Silva-Marcos, J. High Energy Phys. {\bf 0212}, 036 (2002);
                         S. Sultansoy  {\em et al.},  Acta Phys. Polon. B{\bf 37}, 2839 (2006).

   \bibitem{gut}         N. Borstnik {\em et al.}, Bled Workshops in Physics, Vol.7, No. 2 (2006).

   \bibitem{pdg}         W. M. Yao  {\em et al.} [Particle Data Group], J. Phys. G {\bf 33}, 1 (2006).

   \bibitem{kribs}       G. Kribs  {\em et al.}, ANL-HEP-PR-07-39 (2007).

   \bibitem{maltoni}     M. Maltoni {\em et al.}, Phys. Lett. B {\bf 476}, 107 (2000).

   \bibitem{astro}       D. Fargion {\em et al.}, JETP Lett. 69, 434 (1999).

   \bibitem{okun}        L. Okun {\em et al.}, Phys. Lett. B {\bf 529}, 111 (2002).

   \bibitem{he}          H.-J. He, N. Polonsky, and S. Su, Phys. Rev. D {\bf 64}, 053004 (2001).

   \bibitem{frampton}    P.Frampton, P. Hung, and M. Sher, Phys. Rept. 330, 263 (2000).

   \bibitem{bm}          D. Choudhury, T. M. P. Tait, and C. E. M. Wagner,
                         Phys. Rev. D {\bf 65}, 053002 (2002).

   \bibitem{chanowitz}   M. Chanowitz, Phys. Rev. Lett. {\bf 87}, 231802 (2001).

   \bibitem{little-higgs} T. Han {\em et al.}, Phys. Lett. B {\bf 563}, 191 (2003); B. A. Dobrescu
                         and C. T. Hill, Phys. Rev. Lett. {\bf 81}, 2634 (1998).

   \bibitem{t-parity}    H. C. Cheng and I. Low, J. High Energy Phys. {\bf 0408}, 61 (2004).

   \bibitem{twof}        D. E. Kaplan and M. Schmaltz, J. High Energy Phys. {\bf 0310}, 39 (2003);
                         R. Barcelo, M. Masip, and M. Moreno-Torres, arXiv:hep-ph0701040. 

   \bibitem{ref-eta}     We use a cylindrical coordinate system about the beam axis in which
                         $\theta$ is the polar angle and $\phi$ is the azimuthal angle.  We define 
                         the pseudorapidity $\eta$ of a particle's three-momentum as 
                         $\eta \equiv -\ln(\tan\frac{\theta}{2})$.  The transverse
                         energy and momentum are defined as $E_T = E \sin\theta$ and $p_T = p\sin\theta$
                         where E is the energy measured by the calorimeter and $p$ is the momentum
                         measured in the tracking system.  The missing transverse energy is defined as
                         $\met = |-\sum_i E_T^i \vec{n}_i |$ where $\vec{n}_i$ is a unit vector in the 
                         transverse plane that points from the event vertex to the $i$th calorimeter tower.

   \bibitem{cdf}         D. Acosta {\em et al.} (CDF Collaboration), Phys. Rev. D {\bf 71}, 032001 (2005).

   \bibitem{ref-lum}     S. Klimenko, J. Konigsberg, and T.M. Liss, FERMILAB-FN-0741 (2003).

   \bibitem{top-kin}     D. Acosta {\em et al.} (CDF Collaboration), Phys. Rev. D {\bf 72}, 052003 (2005).

   \bibitem{jet-energy}  A. Bhatti {\em et al.}, Nucl.~Instrum.~Methods Phys. Res. A {\bf 566}, 375  (2006).

   \bibitem{alpgen}      M.~L.~Mangano {\em et al.}, J. High Energy Phys. {\bf 0307}, 1 (2003).

   \bibitem{herwig}      G. Corcella  {\em et al.}, J. High Energy Phys. {\bf 0101}, 10 (2001).

   \bibitem{pythia}      T. Sj{\" o}strand {\em et al.}, Comput. Phys. Commun. {\bf 135}, 238 (2001),
                         and references therein.

  \bibitem{topmass}      A. Abulencia {\em et al.} (CDF Collaboration) , Phys. Rev. D {\bf 73}, 032003 (2006).

  \bibitem{cacciari}     M. Cacciari {\em et al.}, J. High Energy Phys. {\bf 0404}, 68 (2004).

\end{thebibliography}
\end{document}